\documentclass[oldversion]{aa}
\usepackage{natbib}
\usepackage{epsfig}
\usepackage{xspace}
\usepackage{amssymb}
\usepackage{amsmath}
\usepackage{graphicx}

\def\Sc{\ensuremath{\mathrm{Sc}}\xspace}

\newcommand{\simless}{\mathbin{\lower 3pt\hbox
      {$\rlap{\raise 5pt\hbox{$\char'074$}}\mathchar"7218$}}}
\newcommand{\simgreat}{\mathbin{\lower 3pt\hbox
     {$\rlap{\raise 5pt\hbox{$\char'076$}}\mathchar"7218$}}}

\begin{document}
\title{Dust crystallinity in protoplanetary disks: 
the effect of diffusion/viscosity ratio}
\titlerunning{Dust crystallinity in protoplanetary disks}
\author{Ya.~Pavlyuchenkov \and C.~P. Dullemond}
\offprints{Ya.~Pavlyuchenkov, \email{pavyar@mpia.de}}
\institute{Max Planck Institut f\"ur Astronomie, K\"onigstuhl 17, 69117
Heidelberg, Germany}

\abstract{
The process of turbulent radial mixing in protoplanetary disks has
strong relevance to the analysis of the spatial distribution of
crystalline dust species in disks around young stars and to studies of
the composition of meteorites and comets in our own solar system. 
 A debate has gone on in the recent literature on the ratio of the
effective viscosity coefficient $\nu$ (responsible for accretion)
to the turbulent diffusion coefficient $D$ (responsible for mixing).
Numerical magneto-hydrodynamic simulations have yielded values between
$\nu/D\simeq 10$ (Carballido, Stone \& Pringle \citeyear{carballido:2005})
and $\nu/D\simeq 0.85$ (Johansen \& Klahr \citeyear{johansenklahr:2005}).
Here we present two analytic arguments for the ratio $\nu/D=1/3$ which
are based on elegant, though strongly simplified assumptions. We argue
that whichever of these numbers comes closest to reality may be determined
{\em observationally} by using spatially resolved mid-infrared
measurements of protoplanetary disks around Herbig stars. If meridional
flows are present in the disk, then we expect less abundance of
crystalline dust in the surface layers, a prediction which can likewise
be observationally tested with mid-infrared interferometers.
}
\maketitle

\section{Introduction}
Turbulent mixing plays an important role in the physics of
protoplanetary accretion disks. The same turbulence that is responsible
for the anomalous viscosity of the disk (and thus for the accretion
process) is also responsible for the radial and vertical mixing of
material in the disk. This mixing is likely to have strong influence on
thermochemical processes and grain growth in disks. For instance, models
of the gas-phase chemistry in disks turn out to be strongly dependent on
the level of turbulent mixing (Gammie \citeyear{gammie:1996}; 
Semenov, Wiebe \& Henning \citeyear{semenovwiebe:2004}; Ilgner
\citeyear{ilgnernelson:2006a}).  This is because some chemical
reactions may be very slow in some regions but fast in another. Thus
mixing can transport processed gas from the latter region  to the former
region, affecting the molecular abundances there.  

For dust solid-state chemistry mixing plays presumably an even bigger role.
Some evidence from meteoritics points to weak mixing, such as the recently
discovered chemical complementarity of chondrules and matrix in some chondrites
(Klerner \& Palme \citeyear{klernerpalme:2000}). On the other hand, the
existence of Calcium-Aluminium-rich Inclusions (CAIs) in many chondritic
meteorites points toward some level of mixing in the early solar system,
since CAIs are thought to form closer to the sun. Samples returned from
comet Wild 2 by the STARDUST mission revealed that CAI-like material
can be present in comets, which is interpreted as evidence for radial
mixing (Zolensky et al.~\citeyear{zolensky:2006}). In infrared
observations of protoplanetary disks there is strong evidence of the
existence of crystalline silicates at radii in the disk that are far
larger than the radius at which the disk temperatures are high enough
for thermal annealing. One idea is that they got there by radial mixing
from these hot inner regions to the cooler outer regions (Gail
\citeyear{gail:2001}; Bockelee-Morvan \citeyear{bockmorvan:2002}).

The process of radial mixing by turbulence can be described by a
diffusion equation governed by a diffusion coefficient $D$ (Morfill \&
V\"olk~\citeyear{morfillvoelk:1984}). Turbulence also plays an important
role in the theory of coagulation of dust (e.g.~V\"olk et
al.~\citeyear{voelkmorroejon:1980}). It can both prevent and accelerate
grain growth, in complex ways. The process of accretion is, on the other
hand, described by the equations of mass- and angular momentum
conservation in a viscous medium, which is ruled by the effective
viscosity coefficient $\nu$ (Shakura \& Sunyaev
\citeyear{shaksuny:1973}; Lynden-Bell \& Pringle
\citeyear{lyndenpring:1974}). Both $\nu$ and $D$ may depend on distance to
the star (i.e.~the radial coordinate of the disk $R$) and both have the
same dimension. In fact, because they both arise from the same
turbulence, there are reasons to believe that they are nearly the same,
apart from a dimensionless factor of order unity which is called the
Schmidt number: $\nu/D\equiv\Sc$. The effective viscosity $\nu$
may be partly produced by Reynold stress (turbulent motions of the gas),
Maxwell-stress (magnetic field lines transporting angular momentum),
as well as gravity waves in moderately gravitationally unstable disks.
However, the mixing $D$ can {\em only} be due to Reynold stress
(turbulent motions). So $D$ and $\nu$ are clearly related, but do not
necessarily have to be equal.

As has been shown by Clarke \& Pringle~(\citeyear{clarkepringle:1988}), the
efficiency of outward radial mixing in steady disks depends on the {\em
ratio} of $\nu$ to $D$, i.e.~on $\Sc$, and therefore the Schmidt number
plays an essential role in the understanding of the distribution of various
chemical and dust species in disks, much more so than the absolute values of
$D$ and $\nu$. The Schmidt number of turbulent flows has been discussed in
several studies in the past (e.g.~Tennekes \& Lumley
\citeyear{tennekeslumley:1972}; McComb \citeyear{mccomb:1990}). These
authors typically argue for $\nu/D\simeq 0.7$. It is not clear, however,
whether these findings can be directly translated to turbulence in rotating
Keplerian accretion disks. Accretion disk flows are fundamentally different
from laboratory flow (Balbus \citeyear{balbusreview:2003}).  
Stone \& Balbus~(\citeyear{StoneBalbus:1996}) have shown, using 3D hydrodynamical
simulations as well as analytical arguments, that hydrodynamical turbulence
resulting from vertical convection tends to transport angular momentum inward
instead of outward. This stands in contrast to turbulence in planar shear 
flows which, as Stone \& Balbus confirm in their simulations, transports 
momentum in the opposite direction. The same kind of inward angular momentum 
transport was reported by R\"udiger et al.~(\citeyear{ruedigeregorov:2005}). 

Currently the most widely accepted theory for the origin of effective viscosity
in accretion disks is that of magneto-rotational turbulence. This picture
is based on an interplay between weak magnetic fields and the
differentially rotating gas in the disk, causing an instability that
drives the turbulence (Balbus \& Hawley~\citeyear{balbushawley:1991}). 
Angular momentum is then transported both by Reynold stresses as well 
as by Maxwell stresses. However, there are various uncertainties about whether
the magneto-rotational instability can operate in protoplanetary disks
which are often very near to being neutral
(e.g.~Semenov et al.~\citeyear{semenovwiebe:2004}; Ilgner \& Nelson~\citeyear{ilgnernelson:2006a},
Oishi et al.~\citeyear{oishi:2007}). All in all it is still quite unclear
what the nature of the turbulence and the mechanism of  angular momentum
transport is, and therefore the issue of the $\nu/D$ ratio also remains open.

Recently, several publications have attempted to shed light on the $\nu/D$
ratio in accretion disks with full 3D MHD simulations of such turbulence.
 These simulations have yielded values ranging from $\nu/D\simeq 10$
(Carballido et al.~\citeyear{carballido:2005}) to $\nu/D\simeq
0.85$ (Johansen \& Klahr~\citeyear{johansenklahr:2005}). The drawback of
these 3D MHD slab geometry models is that they have a finite numerical
resolution which is in general rather coarse. There are reasons to believe
that simulations with higher resolution might yield different results.
Such initial calculations seem to indeed indicate a  reduction of
magneto-rotational instability, (Dzyurkevich 2007, private communication).

At the same time, new 2D models of Keller \& Gail~(\citeyear{keller:2004})
and Tscharnuter \& Gail~(\citeyear{tscharnuter:2007}) show the presence of
large-scale circulations  within the disk. These calculations support
the results found previously in the analytical paper of Urpin~(\citeyear{urpin:1984}),
and in the asymptotic study of Regev \& Gitelman~(\citeyear{regev:2002}).
Following these results, beyond the certain critical radius near the disk's
equatorial  plane, the material moves in the outward direction, whereas the
accretion flow develops in the surface layer of the disk. As a result of
the large-scale circulation, which is driven by viscous angular momentum
transfer, advective transport dominates diffusive mixing in the outer part
of the disk. Species that are produced or undergo chemical reactions in the
warm inner zones of the disks are advectivelly  transported into the
cool outer regions. Such dynamics cannot be purely represented  by any
1D viscous or diffusional models. Thus, $\nu/D$ ratio as direct  measure
of the mixing efficiency is not appropriate in this picture. But,
despite the beauty of this  theory, it is based on the postulated
anomalous $\alpha$-viscosity,  which is an apriory micro-scale quantity.
If strong turbulence exists on the scale which is comparable to 
the height of the disk, it may destroy such a circulation pattern.

The question is now: which scenario of the disk evolution (viscous, diffusion
or advective) is the right one in nature? From theoretical arguments it is
still challenging to tell. But we propose here an observational test that
may possibly distinguish between all these cases. This allows us
to `measure' effective $\nu/D$ for stars as bright as Herbig Ae stars.
Unfortunately this will not work for T Tauri stars and Brown Dwarfs,
due to lack of spatial resolution.

This paper is organized as follows. In Sect.~\ref{sec-deriv-packages}
we re-derive the Pringle (\citeyear{pringle:1981}) equation on the basis of
a simple diffusion recipe, and we thereby obtain the Schmidt number under the
assumption that the simple diffusion recipe describes the true motion of gas
parcels in the disk. In Sect.~\ref{sec-deriv-independent} we derive the
Schmidt number in a different way, by assuming that each component of the
fluid obeys the same equation. We obtain the same Schmidt number as in
Sect.~\ref{sec-deriv-packages}. In Sect.~\ref{sed-abundance-powerlaw} we
review how the Schmidt number affects the abundance of crystalline silicates
in the inner disk regions. In Sect.~\ref{sec-measuring} we present
radiative transfer results for the mid-infrared spectra of spatially
resolved disks, and propose how the study of such spectra might provide
insight into what is the value of the Schmidt number in real protoplanetary
disks.

In this paper we focus purely on the diffusion of gas or of small particles
that are well coupled to the gas. Bigger dust particles (in the size range
of 1 cm or bigger) will decouple from the turbulence and drift inward. This
is another process which we do not include in this paper. Also, for
simplicity we assume that the disk is vertically averaged and axially
symmetric.

\section{The diffusion equation for the Keplerian disk:
derivation from a discrete model}\label{sec-deriv-packages}

In this section we will re-derive the well known equation 
of Pringle~(\citeyear{pringle:1981}) for the spreading of the accretion disk.
The derivation is based on the idea that the evolution of the disk can be
completely described by a special kind of turbulent motion of fluid parcels.
In normal turbulence each eddy of the disk has a deviation $\delta v$ from the
average velocity of the fluid. These are generally non-Keplerian. But the
conservation of angular momentum also prevents these gas parcels from drifting
too far inward or outward, unless they exchange angular momentum with
neighboring gas parcels, which leads to angular momentum transfer in the wrong
direction (see above). However, it is known that weakly magnetized disks are
unstable to the magneto-rotational instability 
(Balbus \& Hawley~\citeyear{balbushawley:1991}). Magnetic field
lines enable two parcels at different radii to exchange angular momentum
`over a distance', if they are thread by the same field lines. We suggest
an extremely simplified picture for the motions of these parcels: Consider
two parcels, one slowly drifting inward, the other slowly drifting
outward. We assume that they both have a locally Keplerian angular velocity
at all times. This means that the inward moving parcel loses angular
momentum while the other gains angular momentum. We assume that neither
parcel has angular momentum exchange with its surroundings. So angular
momentum can only be conserved if the loss of angular momentum of the inward
moving parcel is compensated by the gain of angular momentum by the outward
moving parcel.  We invoke an unspecified long-range force that enables this
exchange of angular momentum between the two parcels. Physically, the only
long-range force that might facilitate this is the Lorenz force, by magnetic
fields, so we assume that magnetic fields indeed provide this torque. 
The accretion and spreading of the disk is, in this picture,
nothing other than a material diffusion: Packages of gas being transferred
randomly, under the condition of local angular momentum conservation. The
radial mixing of a passive tracer in the disk is therefore an integral part
of this. If the accretion/spreading of the disk is then viewed in a
framework of `viscous disks', then the $\nu/D$ ratio follows strictly from
this derivation. 

This picture was also presented in 
Tutukov \& Pavlyuchenkov~(\citeyear{tutukovPav:2004}), where different
types of astrophysical disks are described in frames of non-stationary
diffusional models. However, they used only numerical models and 
did not provide the equation which governs the evolution of such 
disks.

In this section we will derive these equations and cast them in the
standard form of viscous disk equations, which yields the ratio $\nu/D$. We
are fully aware that the simplifications of our assumptions are quite
drastic and that realistic MHD simulations are presumably better and yield
somewhat different results.  Nevertheless, we justify our approach in
light of the fundamental nature of these equations.

\subsection{The mass fluxes from the cell}
Let the mass of the disk be small enough so its self-gravitation
can be neglected. The disk is assumed to be geometrically thin,
Keplerian, and axially symmetric.  We describe the structure of the
disk in terms of the surface density $\Sigma (R)$, where $R$ is the 
distance from the axis of symmetry.  

We introduce a grid \{$R_i$\}, which divides the domain into annular
elements which form the grid cells. We assume that the cell size
$h_i$=($R_{i+1}$-$R_{i}$) is equal to the {\it characteristic radial mean
free path of a turbulent element}. The surface density $\Sigma_i$ is assumed
to be constant within each cell. Along with $\Sigma_i$, the average orbital
Keplerian velocity is also determined in each grid cell, $V_i$, as well as
the average radial component of the turbulent velocity,
$V_{r,\mathrm{turb}}(i)$.

\begin{figure}[b]
\centering
\includegraphics[width=0.45\textwidth]{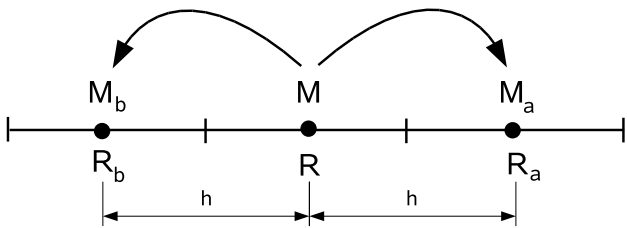}
\caption{The scheme of the mass transfer in the model
of diffusion disk.} 
\label{scheme1}
\vspace{0.5cm}
\end{figure}

Let us consider a current cell and denote the nearest cells A and B, see
Fig.~\ref{scheme1}. 
We suppose that in a time $\Delta t$ the mass $M$ in the
current cell moves to the neighboring cells due to turbulent motion. The
matter leaving this cell is transferred to the two adjacent cells, and this
redistribution of the mass must conserve mass and angular momentum according
to the philosophy described above. The corresponding equations for the
system of three cells can be written
\begin{align}
&M = M_a+M_b \label{eq1} \\
&M R V = M_a R_a V_a + M_b R_b V_b, \label{eq2}
\end{align}
where $V_a$ and $V_b$ are the Keplerian velocities for radii
$R_a$ and $R_b$, respectively.

Here we will derive the differential equation which describes the
evolution  of such a disk in the limit $h \rightarrow 0$, assuming also
for simplicity  that $h_i$=$h$=const.
Using the Taylor expansion of Eq.~\eqref{eq2} we obtain 
(see Appendix~A1):
\begin{align}
&M_a = \dfrac{M}{2}\left(1+\dfrac{1}{4}\dfrac{h}{R}\right) \label{eq6} \\
&M_b = \dfrac{M}{2}\left(1-\dfrac{1}{4}\dfrac{h}{R}\right). \label{eq7}
\end{align}
Note that this looks as if mass is always transported outward,
since $M_b < M_a$. However, as we shall see below, the true radial
flux of matter may have either sign.

\subsection{The mass flux across the cell boundary}
\begin{figure}
\centering
\includegraphics[width=0.3\textwidth]{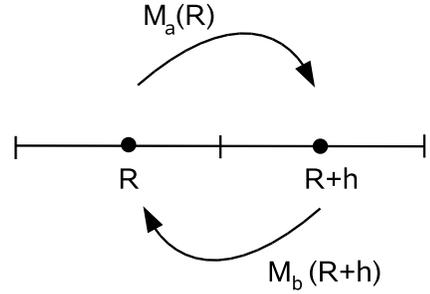}
\caption{The scheme of the mass transfer between two cells
in the model of diffusion disk} 
\label{scheme2}
\vspace{0.5cm}
\end{figure}

Denote $M_a^{(R)}$ as a mass transferred from the cell with $R$ into the
cell $R+h$. Also, let $M_b^{(R+h)}$ be a mass transferred from the cell
$R+h$ into the cell $R$, see  Fig.~\ref{scheme2}.
Denote also $F_{ab}$ as a total flux across the boundary
between the cells:
\begin{equation}
F_{ab}\Delta t=M_a^{(R)}-M_b^{(R+h)}. \label{eq10}
\end{equation}
Using the Taylor expansion of Eq.~\eqref{eq10} (see Appendix~A2) we find: 
\begin{equation}
F_{ab} \Delta t = - \dfrac{1}{2}h R^{1/2} \dfrac{\partial}{\partial R}(MR^{-1/2}). 
\label{eq12}
\end{equation}
Note, that if the mass flux $M^{(R)}$ does not depend on radius, then 
$F_{ab}>0$, i.e. the mass is transferred outwards.

One of the basic assumptions of our diffusion model is the relation 
for the flux from the cell:
\begin{equation}
M = 2\pi R \cdot \Sigma \cdot V_{r,\mathrm{turb}} \cdot \Delta t. 
\label{eq13}
\end{equation}
If we substitute this relation into Eq.~\eqref{eq12} and introduce 
the diffusion coefficient (see Appendix~A3 for the choice of numerical
coefficient):
\begin{equation}
D= \frac{1}{2} h \cdot V_{r,\mathrm{turb}},
\label{eq14}
\end{equation}
we get the final formula for the total flux across the cell boundary: 
\begin{equation}
F_{ab} = -\pi R^{1/2} \dfrac{\partial}{\partial R}(D\Sigma R^{1/2}). 
\label{eq15}
\end{equation}
Note now that if $\Sigma$ and $D$ do not depend on radius than $F_{ab}=-\pi
D \Sigma <0$ i.e. the disk becomes an {\em accretion} disk.

\subsection{The equation of the diffusion disk evolution}
The law of the mass conservation means:
\begin{equation}
2\pi R \Delta \Sigma = - \dfrac{\partial F_{ab}}{\partial R}\Delta t,
\label{eq17}
\end{equation}
Using Eq.~\eqref{eq15} we find finally: 
\begin{equation}\label{eq-lbp-diff}
\frac{\partial\Sigma}{\partial t} = \frac{1}{R}\frac{\partial }{\partial R}
\left[\sqrt{R}\frac{\partial }{\partial R}\left(\Sigma D \sqrt{R}\right)\right].
\end{equation}

This is the required equation which describes the evolution of the diffusive 
disk.

If we take $\nu=D/3$ then we can cast Eq.~(\ref{eq-lbp-diff}) into the
familiar equation for accretion disk evolution (see Pringle
\citeyear{pringle:1981}):
\begin{equation}\label{eq-fullsigma-evol}
\frac{\partial\Sigma}{\partial t} = \frac{3}{R}\frac{\partial }{\partial R}
\left[\sqrt{R}\frac{\partial }{\partial R}\left(\Sigma \nu \sqrt{R}\right)\right]
\end{equation}
This equation can be derived from the Navier-Stokes equations for viscous
disks, where $\nu$ is the standard viscosity. Since $D$ is the true
diffusion coefficient of the matter, we have found that the ratio $\nu/D$
must be 1/3 in our picture.

\section{The diffusion equation for Keplerian disk: multicomponent
medium}\label{sec-deriv-independent}
Another way to `derive' the ratio of $\nu/D$ is to start from the {\em
Ansatz} that each gas or dust species evolves independently according to:
\begin{equation}\label{eq-each-species-indep}
\frac{\partial\Sigma_i}{\partial t} = \frac{3}{R}\frac{\partial }{\partial
R} \left[\sqrt{R}\frac{\partial }{\partial R}\left(\Sigma_i \nu
\sqrt{R}\right)\right]
\end{equation}
where $\Sigma_i$ is the surface density of species $i$. The $\Sigma_i$
obey completeness:
\begin{equation}
\Sigma_1 + \Sigma_2 + \cdots + \Sigma_n = \Sigma
\end{equation}
where $n$ is the total number of species. 

The underlying argument to support the Ansatz of
Eq.~(\ref{eq-each-species-indep}) is the following: Suppose that gas+dust
parcels indeed split up and move inward/outward, as assumed in
Sect.~\ref{sec-deriv-packages}. The relative abundances of the various species are
the same in the inward moving parcel as in the outward moving parcel (they
were both from the same original parcel). This means that if the angular
momentum stays constant between the inward and outward moving parcel, the
same can be said of the individual species. So: species $i$ in the inward
moving parcel loses as much angular momentum as the same species in the
outward moving parcel gains. Therefore the diffusion happens in each species
identically, and therefore we argue that Eq.~(\ref{eq-each-species-indep})
holds for each species separately, independent of the other species. This is
even true if one of the species represents 99.9\% of the mass and the other
only $0.1\%$.

So, starting from Eq.~(\ref{eq-each-species-indep}) we can rewrite:
\begin{equation}\label{eq-separ-step1}
\begin{split}
\frac{\partial\Sigma_i}{\partial t} &= \frac{3}{R}\frac{\partial }{\partial
R} \left[\sqrt{R}\frac{\partial }{\partial R}\left(\frac{\Sigma_i}{\Sigma} 
\Sigma \nu \sqrt{R}\right)\right]\\
&=\frac{3}{R}\frac{\partial }{\partial
R} \left[\sqrt{R}\frac{\Sigma_i}{\Sigma}\frac{\partial }{\partial R}\left(
\Sigma \nu \sqrt{R}\right)
+R\Sigma \nu \frac{\partial }{\partial R}\left(
\frac{\Sigma_i}{\Sigma}\right)
\right]
\end{split}
\end{equation}
If we write Eq.~(\ref{eq-fullsigma-evol}) as a conservation equation as
$\dot \Sigma + (1/R)\partial [R\Sigma v_R]/\partial R=0$ where $v_R$ is the
average radial velocity of the matter (including all species), then by
Eq.~(\ref{eq-fullsigma-evol}) this
velocity is given as:
\begin{equation}\label{eq-velo}
v_R = -\frac{3}{\Sigma\sqrt{R}}\frac{\partial}{\partial R}
\left(\Sigma\nu\sqrt{R}\right)
\end{equation}
Using Eq.~(\ref{eq-velo}) in Eq.~(\ref{eq-separ-step1}) one obtains:
\begin{equation}\label{eq-form-morfill}
\frac{\partial\Sigma_i}{\partial t} + \frac{1}{R}\frac{\partial}{\partial R}
\left(R\Sigma_i v_R\right) = \frac{3}{R}\frac{\partial}{\partial R}
\left[R\nu \Sigma\frac{\partial}{\partial R}
\left(\frac{\Sigma_i}{\Sigma}\right)\right]
\end{equation}
This is the same as the equation introduced by Morfill \& V\"olk~(\citeyear{morfillvoelk:1984})
but now with:
\begin{equation}
\frac{\nu}{D} = 1/3
\end{equation} 

So we have shown here that, provided $\nu/D$=1/3, the picture proposed by
Morfill \& V\"olk~(\citeyear{morfillvoelk:1984}) that diffusion takes place
in the comoving frame of an otherwise laminar accretion disk is equivalent
to our picture of exchanging Keplerian parcels. 

Again we stress that our derivation is based on a simplifying assumption.
One can consider our value of $\nu/D=3$ as a minimum value of $\nu/D$. 
At the other extreme is the picture of a perfectly laminar but
viscous disk with $\nu/D=\inf$. The true value must be within these 
limits.

\section{Effect of $\nu/D$ on abundance of crystalline silicates in disks}
\label{sed-abundance-powerlaw}
Although our simple analytic argumentation predicts that $\Sc=1/3$,
MHD simulations and experiments with rotating turbulent flows
show other ratios. 
At the moment we need to consider values ranging from $\Sc=1/3$ (this study)
to $\Sc=10$ (Carballido, Stone \& Pringle~\citeyear{carballido:2005}; Johansen,
Klahr \& Mee~\citeyear{johansenklahrmee:2006}).  So how do these various
values affect the distribution of molecular species and dust species in
the disk?  What are the consequences of this, for instance,
for the problem of crystalline silicates in disks?  One may argue that the
difference between $\nu/D$=1/3 and $\nu/D=1$ is `only' a factor of 3 and
therefore not very strong: the time scales of diffusion will be
different by only a factor of 3. However, for accretion disks in semi-steady
state, a difference of 3 in the diffusion constant can have very strong
effects on the distribution of thermally processed (crystalline) particles
in the disk. If we assume that particles are only thermally processed in the
very inner regions of the disk (i.e.~close to the star) then to find
particles at larger radii, they have to diffuse outward (e.g.~Gail
\citeyear{gail:2001}; Bockelee-Morvan \citeyear{bockmorvan:2002}). The
efficiency of this outward diffusion is very dependent on the $\nu/D$ ratio.
This is due to the fact that the trace particles (crystalline dust) have to `swim
upstream' against the accretion flow to reach larger radii. This was modeled
by Clarke \& Pringle~(\citeyear{clarkepringle:1988}) who showed that the
abundance of such a tracer as a function of radius can be derived
analytically for steady accretion disks. Since in the absence of disk
instabilities the inner regions of viscously evolving disks change over a
time scale much larger than the local viscous time, these inner regions
(say, inward of a few tens of AU at an age of 1 Myr) can be regarded as
being in semi-steady state. This simplifies the analysis drastically. We can
then start from the standard advection-diffusion equation, after Morfill \&
V\"olk~(\citeyear{morfillvoelk:1984}):
\begin{equation}\label{eq-morfill}
\frac{\partial\Sigma_i}{\partial t} + \frac{1}{R}\frac{\partial}{\partial R}
\left(R\Sigma_i v_R\right) = \frac{1}{R}\frac{\partial}{\partial R}
\left[RD\Sigma\frac{\partial}{\partial R}
\left(\frac{\Sigma_i}{\Sigma}\right)\right]
\end{equation}
For the special case of $\nu/D$=1/3 this becomes equal to
Eq.~(\ref{eq-form-morfill}), but we will retain the more general form of
Eq.~(\ref{eq-morfill}) here. For a steady state disk we have (for $R\gg
R_{\mathrm{in}}$ where $R_{\mathrm{in}}$ is the inner radius of the disk):
\begin{equation}\label{eq-steadystate1}
\dot M = - 2\pi R\Sigma v_R,
\end{equation}
with
\begin{equation}\label{eq-steadystate2}
v_R = -\frac{3}{2}\frac{\nu}{R}.
\end{equation}
Here $\dot M$ is the accretion rate of the disk. With the Schmidt number
$\Sc\equiv \nu/D$, by using Eqs.~(\ref{eq-steadystate1}-\ref{eq-steadystate2}), 
and by taking $\partial\Sigma_i/\partial t=0$ (steady state) we can rewrite
Eq.~(\ref{eq-morfill}) in the form:
\begin{equation}
\frac{\partial}{\partial\lg R}\left(\frac{\Sigma_i}{\Sigma}\right)
=-\frac{2}{3}\frac{1}{\Sc}\frac{\partial^2}{\partial(\lg
  R)^2}\left(\frac{\Sigma_i}{\Sigma}\right)
\end{equation}
The solution is:
\begin{equation}\label{eq-mix-sol1}
\frac{\Sigma_i}{\Sigma} = \sigma_i^{(0)}
+ \sigma_i^{(1)} \left(\frac{R_{1}}{R}\right)^{\frac{3}{2}\Sc}
\end{equation}
This is a general steady-state solution for mixing of any tracer in the disk 
for $R\gg R_{\mathrm{in}}$.  The $\sigma_i^{(0)}$ is a background abundance while 
$\sigma_i^{(1)}$ is the abundance at some radius $R_1$.

In general, the Eq.~\eqref{eq-mix-sol1} is appropriate  only if no
large-scale radial flows are present in the disk.  However, it is
possible to modify  this equation in a way that it approximately
accounts for the effect of large-scale advective flows onto emergent spectra.
Following the results of Keller \& Gail~(\citeyear{keller:2004}) and
Tscharnuter \& Gail~(\citeyear{tscharnuter:2007}) 
we assume that near the disk's midplane  the material moves outward, 
while in the surface layers the material flows inward.  
Since we assume that the surface layer of the disk is responsible for 
the emergent spectra we change the Eq.~\eqref{eq-mix-sol1} 
so as it describes the abundances of the tracers in the surface 
layer. Therefore, we substitute $V_R(1+C)$ for $V_R$ in the 
Eq.~\eqref{eq-steadystate1}, where $C$ is the circulation velocity in 
units of the vertically averaged accretion velocity, and thus we
come to a more general relation for the tracer abundances in the surface layer:
\begin{equation}\label{eq-mix-sol2}
\frac{\Sigma_i}{\Sigma} = \sigma_i^{(0)}
+ \sigma_i^{(1)} \left(\frac{R_{1}}{R}\right)^{\frac{3}{2}\Sc\,(1+C)}
\end{equation}
We suppose that $C$ can easily be 2-3 or more which has major 
consequences for the mixing. Given this modification, we introduce
the effective Schmidt number:
\begin{equation}
\bar{\Sc}=\Sc(1+C). 
\end{equation}
Note, that large values  of $C$ `mimic'\, large Schmidt numbers, $\Sc$,
in the picture with  no circulation.

\begin{figure}
\centering
\includegraphics[width=0.45\textwidth]{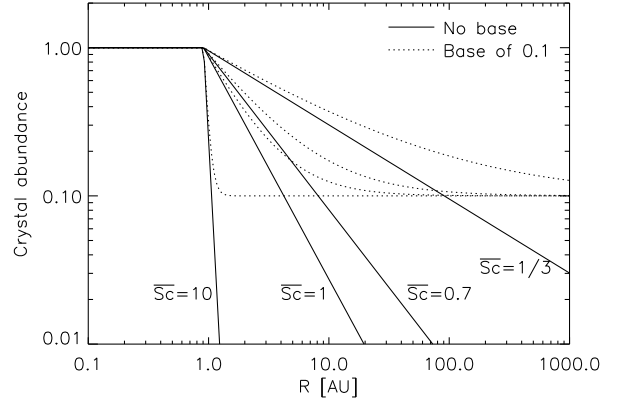}
\caption{\label{fig-abundance}The abundance of crystalline silicates in a
steadily accreting protoplanetary disk around a Herbig star of $M_{*}=2.5
M_{\odot}$, $R_{*}=2.5 L_{\odot}$ and $T_{*}=10^4$K. Solid lines: assuming
that {\em only} thermal annealing in the warm inner regions (here: inward of
about $1$ AU) can produce crystalline silicates. Dotted lines: assuming that
some other mechanism has produced a global base level of crystallinity of
10\%.}
\end{figure}

Let us now apply these results to the problem of crystallinity of dust in a
disk which is hot enough for thermal annealing inward of
$R_{\mathrm{anneal}}$. Since the dependence of the efficiency of thermal
annealing on temperature is extremely sharp, one can say that inward of
$R_{\mathrm{anneal}}$ all dust is crystalline. For $R>R_{\mathrm{anneal}}$
there is no annealing and the mixing solution Eq.~(\ref{eq-mix-sol2}) is
valid in which $R_1\equiv R_{\mathrm{anneal}}$ and
$\sigma_{\mathrm{cryst}}^{(1)}=1$. The background crystallinity
$\sigma_{\mathrm{cryst}}^{(0)}$ is the globally present level of
crystallinity, discussed below. One sees that the abundance of crystalline
silicates at any given radius depends only on $R_{\mathrm{anneal}}$, the
base level of crystallinity $\sigma_{\mathrm{cryst}}^{(0)}$ and the value of
{$\bar{\Sc}$}.  It does {\em not} depend on the value of $\nu$ itself
(i.e.~not on the value of $\alpha$). For strong or weak turbulence the
results are the same, as long as $\bar{\Sc}$ remains the same. 
The solutions to Eq.~(\ref{eq-mix-sol2}) 
for $\sigma_{\mathrm{cryst}}^{(0)}=0$ (no global crystallinity) and 
for $\sigma_{\mathrm{cryst}}^{(0)}=0.1$ (10\% global crystallinity) 
are shown in Fig.~\ref{fig-abundance}.

It is important to keep in mind that these analytic solutions for the level
of crystallinity are only valid for $R_1\equiv R_{\mathrm{anneal}}\gg
R_{\mathrm{in}}$. This is the case for Herbig stars, since they are so
bright that the annealing radius is around 1 AU while the inner disk radius
is more than hundred times smaller. For T Tauri stars, on the
other hand, the annealing radius is only about ten times larger than the
inner radius, which means that the effects of the no-friction boundary
condition of the disk at $R_{\mathrm{in}}$ are still strongly affecting the
solution, and hence the above analytic solution is not completely correct.
Nonetheless, the rough form remains correct even in that case.

Now let us focus on the case without a base level of crystallinity. It
can be seen that the abundance of crystals at large radii is extremely
sensitive to the power index $\bar{\Sc}$. For $\bar{\Sc}=10$, i.e.
for large Schmidt numbers or large surface inflow velocity, there is 
essentially no outward mixing.  For $\bar{\Sc}=1$ or $\bar{\Sc}=0.7$
the crystals are mixed out to a level  of 3\% and 10\% respectively at
10 AU. For $\bar{\Sc}=1/3$ the 10\% crystallinity is reached at 100 AU, i.e.~out
to a hundred times larger radius. This shows that the seemingly small
differences between the various theoretically derived values of $\bar{\Sc}$ make
a very large difference in the distribution of crystalline silicates in
the disk.

So far we have assumed that crystalline silicates are only produced by
thermal annealing in the inner disk regions ($R\le R_{\mathrm{anneal}}$).
However, there is an on going debate about what causes dust grains to become
crystalline. While the thermal annealing and radial mixing of dust (Gail
\citeyear{gail:2001}) is a natural phenomenon that is likely to happen in
all disks, there may be additional mechanisms of crystallizing dust such as
shock heating (Harker \& Desch~\citeyear{harkerdesch:2002}) or lightning
(Gibbard~\citeyear{gibbardlevymorfill:1997} and references
therein). Evidence for these alternative mechanisms is found in infrared
spectroscopy of protoplanetary disks, where it is found in many sources that
the crystalline grains are dominated by forsterite in the outer regions of
many disks. This is contrary to the radial mixing models which predict
that enstatite dominates in the outer disks (Bouwman et al.~in prep.).

From the solar system there is also some evidence that other mechanisms
might crystallize dust or dust aggregates (e.g.~review by Trieloff \&
Palme~\citeyear{trieloffpalme:2006}). However, the inner regions of disks,
near the dust sublimation radius, are always warm enough to assure that
thermal annealing takes place. Therefore, independent of the existence of
other crystallization processes, we expect with near certainty that inward
of some radius all the dust is 100\% crystalline
(i.e.~$\sigma_{\mathrm{cryst}}^{(0)}+\sigma_{\mathrm{cryst}}^{(1)}=1$). The
other crystallization processes, assuming that they can crystallize dust
throughout the disk, produce a base level $\sigma_{\mathrm{cryst}}^{(0)}$ of
crystallinity. The dotted lines in Fig.~\ref{fig-abundance} show how the
combined effects of crystallization work out. We assume here that the time
scales of these additional crystallizing processes are longer than the local
viscous time in the region we study here, so that the level of crystallinity
in the outer disk regions has been built up over a long time and remain
nearly unchanged on the shorter time scales in this inner region.

Another source of a `base-level' of crystallinity
$\sigma_{\mathrm{cryst}}^{(0)}$ is that thermal annealing may also happen
throughout the disk in the very early disk-formation phases (Dullemond, Apai
\& Walch~\citeyear{dullapaiwalch:2006}). They showed that in the stage that
the disk is formed, depending on initial conditions, much of the outward
transport of material may not be due to the `upstream' diffusion, but due to
the fact that the disk material itself is spreading outward (Lynden-Bell \&
Pringle~\citeyear{lyndenpring:1974}). Later, this pre-processed material
accretes back inward and forms a base level of crystallinity
$\sigma_{\mathrm{cryst}}^{(0)}$
(see Dullemond et al.~\citeyear{dullapaiwalch:2006}, Fig.~2).

If large-scale circulations are present in the disk then over a 
large distance there may be vertical mixing between the outward moving 
midplane and inward moving surface layers. This is likely to
produce an enhanced `base'\, value of crystallinity as it inserts 
crystalline grains into the surface layers at large radii.

The conclusion of this section is that radial mixing can have a strong
effect on the distribution of crystalline dust, which is important for
understanding infrared observations of protoplanetary disks, as well as
for understanding the constituents of meteorites. Radial mixing may also
strongly affect disk chemistry or grain growth. A small difference in
the assumed value of $\Sc$ as well as of $C$ could make a strong 
difference in the results of the models.

\section{`Measuring' 
$\bar{\Sc}$ from disks around Herbig Ae stars} \label{sec-measuring}
It is clear that it is of high importance to understand which is the precise
value of $\bar{\Sc}$. Theories will likely always have
uncertainties. In this section we propose a way to derive the value of 
$\bar{\Sc}$ {\em observationally}. From Fig.~\ref{fig-abundance} it is clear that for
Herbig Ae stars the crystallinity of the dust in the region roughly between
$1$ AU and $10$ AU is strongly affected by the ratio $\bar{\Sc}$, even if there
is a base level $\sigma_{\mathrm{cryst}}^{(0)}$ of crystallinity due to
other processes.  And in the case of Schmidt numbers below 1 the radial
mixing may even affect regions as far as out 100 AU. If we can
observationally measure the crystallinity {\em as a function of radius} then
we may be able to determine both $\bar{\Sc}$ as well as 
$\sigma_{\mathrm{cryst}}^{(0)}$.

\begin{figure*}
\centering
\includegraphics[width=0.8\textwidth]{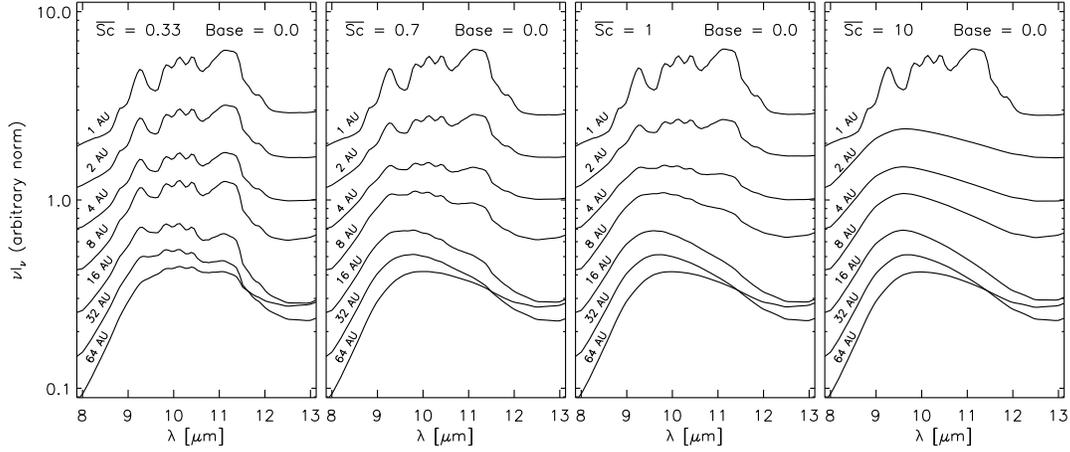}
\caption{Intensity spectra of the disk (arbitrarily normalized) at different
radii, for different assumed $\bar{\Sc}$, for the standard model shown
in Fig.~\ref{fig-abundance}. Here the base level of crystallinity
$\sigma_{\mathrm{cryst}}^{(0)}$ is chosen to be 0, i.e.~the spectra
correspond to the crystallinity levels shown as solid lines in
Fig.~\ref{fig-abundance}.}
\label{fig-intspec-nobase}
\vspace{0.5cm}
\end{figure*}

\begin{figure*}
\centering
\includegraphics[width=0.8\textwidth]{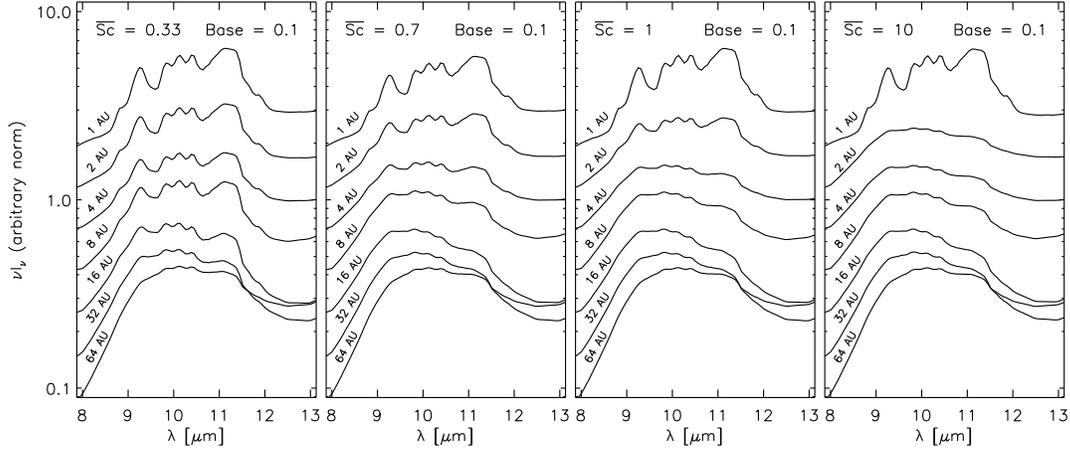}
\caption{Same as Fig.~\ref{fig-intspec-nobase}, but with a base
level of crystallinity of $\sigma_{\mathrm{cryst}}^{(0)}=0.1$.}
\label{fig-intspec-base}
\vspace{0.5cm}
\end{figure*}

The abundance of crystalline dust in a disk can be determined from the
analysis of infrared spectra (see review by Natta et
al.~\citeyear{nattappv:2007}).  For ground-based observations the 8-13
$\mu$m window is well-suited for this. There are by now a number of
instruments and telescopes that can spectrally and spatially resolve disks
in the 10 micron regime at spatial resolutions of about 20 AU for typical
sources: e.g.~VISIR at the VLT, COMICS at Subaru and T-ReCS at
Gemini-South. A spatial resolution of about 1 AU can be obtained with
mid-infrared interferometry with the MIDI instrument on the VLT (Leinert et
al.~\citeyear{leinertvanboekel:2004}).  Although in particular the
interferometry observations do not directly yield spectra as a function of
radius, they do give information on those spatial scales that can be
compared to model predictions. Of course these model predictions have to be
done for each source individually. 

To demonstrate the principle, we show
the intensity-spectrum at various radii in a model of a disk around a
Herbig Ae/Be star of $M_{*}=2.5 M_{\odot}$, $R_{*}=2.5 L_{\odot}$ and
$T_{*}=10^4$K, i.e. the same system as was shown in
Fig.~\ref{fig-abundance}. We took a disk with a mass of 0.0015 $M_{\odot}$,
a size of 90 AU and a surface density profile of $\Sigma(R) \propto R^{-1}$. 
The 2-D axisymmetric density structure of the model was taken to
be $\rho(r,z)=\Sigma(R)\exp(-z^2/2R^2)/(H_p(R)\sqrt{2\pi})$, where $H_p(R)$
is the pressure scale height estimated from the midplane temperature of
a simple flaring disk model of the kind of Chiang \& Goldreich~(\citeyear{cg97}).
We included two dust species, both consisting of 0.1 $\mu$m size grains.  The
first species is a mixture of 25\% amorphous olivine, 25\% amorphous
pyroxene and 50\% carbon.  The second species consists of 25\% crystalline
enstatite and 25\% crystalline forsterite and 50\% carbon. Within each
species the various components are in thermal contact, but the two species
are not in thermal contact with each other. We used optical constants from
Dorschner et al.~(\citeyear{dorschner:1995}) for the amorphous olivine and
pyroxene, from Servoin \& Piriou~(\citeyear{servoin:1973}) for crystalline
forsterite and from J\"ager et al.~(\citeyear{jaeger:1998}) for crystalline
enstatite. The method of computation of the opacities from these constants
is described in Min et al.~(\citeyear{minhove:2005}).
Given the density structure, the opacities and the stellar parameters we
can apply a  Monte-Carlo radiative transfer code
(RADMC, see Dullemond \& Dominik~\citeyear{duldomdisk:2004}) to compute the
dust temperature everywhere in the disk. RADMC is a Monte Carlo code for
dust continuum radiative transfer based on the method of 
Bjorkman \& Wood~(\citeyear{BW:2001}) combined with the method of
Lucy~(\citeyear{Lucy:1999}). This gives the typical structure
of irradiated disks which have a warm surface and a cooler interior.
A ray-tracing program is then applied to compute, from these density and
temperature structures, a spectrum. Due to the warm surface layer lying on
top of a cooler interior, the dust features appear in emission in the
spectrum.
Figure~\ref{fig-intspec-nobase} shows the intensity spectra
($I_\nu(R)$) obtained in this way at
various radii (arbitrarily normalized) for four different values of $\Sc$,
assuming that there is no base level of crystallinity, i.e.\, that thermal
annealing in the inner disk regions is the only source of crystallinity.
The intensity spectrum is the intensity of the face-on disk image as
seen at that particular radius, as a function of wavelength.
In Fig.~\ref{fig-intspec-base} the same is shown, but including a base level
$\sigma_{\mathrm{cryst}}^{(0)}$ of crystallinity due to either the disk's
formation history (e.g.~Dullemond et al.~\citeyear{dullapaiwalch:2006})
or due to additional crystallization processes (e.g.~Harker \& Desch
\citeyear{harkerdesch:2002}).

From Figs.~\ref{fig-intspec-nobase} and \ref{fig-intspec-base} one can see
that both the base level $\sigma_{\mathrm{cryst}}^{(0)}$ and the
$\bar{\Sc}$ affect the spectra in a way that would allow the
reconstruction of these two values from the observations. The
$\sigma_{\mathrm{cryst}}^{(0)}$ is reflected in the fact that the strength
of the crystalline features decreases with radius but gets `stuck' at some
level which is set by the value of $\sigma_{\mathrm{cryst}}^{(0)}$. The
$\bar{\Sc}$ value is reflected in the rapidity by which the crystalline features
become weaker with radius as one goes to larger radii. And
$R_{\mathrm{anneal}}$ is found as the radius by which the crystallinity
suddenly starts to drop. In principle the $R_{\mathrm{anneal}}$ should
follow directly from theoretical considerations, as the dust temperature in
the disk's surface layer, for a given grain size, can be determined
theoretically. However, the grain size is not known beforehand, and an
uncertainty of the grain size in the $0.1$ to $1$ $\mu$m regime can not be
constrained by the analysis of the $10$ $\mu$m feature shape, but it does
influence the dust temperature.  Therefore an observational
determination of $R_{\mathrm{anneal}}$ is necessary and possible.

However, there are some difficulties with this scheme. One obvious problem
is that one requires rather accurate measurements of the spectrum at many
radii. Since most of the action takes place very close to 1 AU, these
measurements would rely strongly on the MIDI interferometer at the VLT or its
planned successor, and it would require measurements with at least 4
baselines that are all ideally aligned in the same position angle on the
sky. Moreover it would be best if the disk is nearly face-on and bright.

Another complication is that thermal annealing may take place in the surface
layers at a different radius than the midplane. If the disk's heating is
dominated by irradiation, then the midplane temperature of the disk is much
smaller than the surface temperature. In this case there are radii where the
dust is above annealing temperature in the surface, but below annealing
temperature in the disk's midplane. It then depends strongly on the {\em
vertical} mixing efficiency whether the annealing in the surface can also
strongly affect the crystallinity of dust in the midplane. If not, then the
$R_{\mathrm{anneal}}$ relevant for the radial mixing might be smaller than
the radius where the surface is annealed. One would then expect, as a
function of radius, the crystallinity to be unity out to that surface
annealing radius and then to suddenly drop to the level of the power-law
which originated at a smaller annealing radius. In principle this could also
be observed, if sufficiently finely-spaced measurements are taken. If
accretional (viscous) heating could heat the disk's midplane sufficiently
strongly, then the oppose may be true: the annealing radius would then be
larger than the annealing radius computed with irradiation only, and only
drop at larger radii. Moreover, if meridional flows are present, the 
picture might blur even more.

It is clear that the full interpretation of these measurements will not be
easy.  But since the Schmidt number is such a fundamental number for
accretion disk theory, it might be worthwhile to attempt to do such
measurements nonetheless.

\section{Conclusions}

In this paper we discuss the ratio between  turbulent  diffusion
and viscosity in protoplanetary disks and study an
effect of this ratio on  the dust crystallinity. Our conclusions can be
summarized as follows:

\begin{enumerate}

\item The efficiency of radial mixing of any tracers in a steady 
state disk depends on the ratio between viscosity and diffusion 
coefficients i.e. on the Schmidt number $\Sc=\nu/D$ if we assume that 
the disk is well mixed in vertical direction. Therefore, the Schmidt 
number plays an essential role in the understanding of the distribution 
of the various chemical and dust species in disks.

\item We present the simple analytic model of the diffusively spreading 
disk. We show that evolution of the diffusively spreading disk 
and evolution of the viscousely spreading disk are described by the 
same equations if we assume $\nu/D=1/3$ where $\nu$ is the  effective
viscosity  coefficient in the classical Shakura \& Sunyaev~(\citeyear{shaksuny:1973}) 
picture. This does not mean, however, that we necessarily argue 
that true accretion disks have this $\Sc=1/3$: this depends on the validity 
of our assumptions  (vertically averaged disk, good mixing between 
dust and gas, a particular kind of turbulence). However, one can regard the
value $\Sc=1/3$ as the lower limit of the Schmidt number which can be included
in the models of radial mixing in Keplerian disks.

\item If large-scale meridional flows are present in the disks then the
Schmidt number cannot be used as a direct measure of the mixing efficiency.
However, it is possible to approximately describe the abundance of the 
tracers in the surface layers of the disk by introducing the 
effective Schmidt number $\bar{\Sc}=\Sc(1+C)$ where $C$ is the 
layer velocity in units of vertically averaged accretion velocity. This
approximation is valid as long as no vertical mixing between 
in- and outward moving layers is assumed. In this case, we expect less 
abundance of crystalline dust in the surface layers.

\item We show that the value of $\bar{\Sc}$ may possibly be measured 
in protoplanetary disks around young stars by using mid-infrared 
interferometry. This requires measurements of a single source at many 
baselines which all lie along the same position angle across the sky.

\end{enumerate}

\begin{acknowledgements}
We are grateful to E.~Kurbatov and A.V.~Tutukov for help in derivation
of the diffusion equation. We are grateful to N.~Dzyurkevich for
a careful review of the paper. We also wish to thank R.~van Boekel,
J.~Bouwman and M.~Min for useful discussions and for the opacity
tables. We thank the anonymous referee for very insightful comments
which have significantly improved the paper and reminded us of the
importance of circulation in disks.
\end{acknowledgements}

\section*{Appendix A1} 
Consider Eq.~\eqref{eq2}, and produce the Taylor expansion around central cell, $R$:
\begin{eqnarray}
M R V = M_a (R+h)\left(V+\dfrac{\partial V}{\partial R} h + \dfrac{1}{2} \dfrac{\partial^{2}V}{\partial R^{2}} h^{2}\right) \nonumber \\ 
+ M_b (R-h)\left(V-\dfrac{\partial V}{\partial R} h + \dfrac{1}{2} \dfrac{\partial^{2}V}{\partial R^{2}} h^{2}\right). \label{eq3a}
\end{eqnarray}
After elementary operations, using Eq.~\eqref{eq1} and dropping terms
of order $h^3$ we get:
\begin{equation}
(M_b-M_a) = \dfrac{Mh}{2}\frac{\dfrac{\partial^2}{\partial R^2}\left(RV\right)}
{\dfrac{\partial}{\partial R}\left(RV\right)}. \label{eq4a}
\end{equation}
For the Keplerian disk, $V \sim R^{-1/2}$, this relation becomes:
\begin{equation}
(M_b-M_a) = -\dfrac{M}{4}\dfrac{h}{R}. \label{eq5a}
\end{equation}
Using Eq.~\eqref{eq1} we obtain:
\begin{align}
&M_a = \dfrac{M}{2}\left(1+\dfrac{1}{4}\dfrac{h}{R}\right) \label{eq6a} \\
&M_b = \dfrac{M}{2}\left(1-\dfrac{1}{4}\dfrac{h}{R}\right). \label{eq7a}
\end{align}

\section*{Appendix A2}
Rewrite the fluxes $M_a^{(R)}$ and $M_b^{(R+h)}$ using Eq.~\eqref{eq6} and~\eqref{eq7}:
\begin{align}
&M_a^{(R)} = \dfrac{1}{2}M^{(R)}\left\{1+\dfrac{1}{4}\dfrac{h}{R}\right\} \label{eq8a} \\
&M_b^{(R+h)} = \dfrac{1}{2}M^{(R+h)}\left\{1-\dfrac{1}{4}\dfrac{h}{(R+h)}\right\}. \label{eq9a}
\end{align}
where $M^{(R)}$ and $M^{(R+h)}$ are the masses emerging from the cells
with radii $R$ and $R+h$ respectively.
Denote $F_{ab}$ as a total flux across the boundary between the cells:
\begin{equation}
F_{ab}\Delta t=M_a^{(R)}-M_b^{(R+h)}. \label{eq10a}
\end{equation}
If we take the Taylor expansion of Eq.~\eqref{eq9a} around $R$ 
and leave only the linear terms in $h$ we get:
\begin{equation}
F_{ab} \Delta t = \dfrac{h}{2}\left(\dfrac{1}{2}\dfrac{M}{R}-\dfrac{\partial M}{\partial R}\right). 
\label{eq11a}
\end{equation}
It is convenient to rewrite the last equation in the form: 
\begin{equation}
F_{ab} \Delta t = - \dfrac{1}{2}h R^{1/2} \dfrac{\partial}{\partial R}(MR^{-1/2}). 
\label{eq12a}
\end{equation}

\section*{Appendix A3} 
Let us comment on an important point in the above derivation. One may argue that 
the factor 1/2 in Eq.~\eqref{eq14} is an arbitrary number. Thus, the factor 
which appears in the right part of final Eq.~\eqref{eq-lbp-diff} 
(now it is 1) is also arbitrary. However, we support our choice 
of 1/2 in Eq.~\eqref{eq14} based on the following arguments. It is easy to rewrite 
the derivation for 1D planar geometry i.e. to obtain the diffusion 
equation in Cartesian coordinates. In terms of the above equations we should simply 
set $M_a=M_b=M/2$ instead of using Eqs.~\eqref{eq6}--\eqref{eq7}. 
The Eq.~\eqref{eq12} is replaced by 
\begin{equation}
F_{ab} \Delta t = -\dfrac{h}{2}\dfrac{\partial M}{\partial R}, 
\end{equation}
and Eq.~\eqref{eq13} is rewritten as
\begin{equation}
M = \Sigma \cdot V_{r,\mathrm{turb}} \cdot \Delta t,
\end{equation}
where $M$ is the total mass of the current cell. 
We should substitute these relations into analogue of Eq.~\eqref{eq17} 
for planar geometry
\begin{equation}
\Delta R \cdot \Delta \Sigma = - (F_{ab}(R+h)-F_{ab}(R))\Delta t.
\end{equation}
In this way we obtain the classical equation:
\begin{equation}\label{eq19}
\frac{\partial\Sigma}{\partial t} = D \frac{\partial^2\Sigma}{\partial R^2},
\end{equation}
only if we introduce the diffusion coefficient, $D$, in the form of Eq.~\eqref{eq14}.
In other words, we choose the convention Eq.~\eqref{eq14} in such a way that 
Eq.~\eqref{eq-lbp-diff} turns into Eq.~\eqref{eq19} in the limit of 
$R \rightarrow \infty$ and $D$=const.

\bibliographystyle{apj.bst}
\bibliography{ms.bbl}

\begin{thebibliography}{32}
\expandafter\ifx\csname natexlab\endcsname\relax\def\natexlab#1{#1}\fi

\bibitem[{Balbus \& Hawley(1991)}]{balbushawley:1991}
Balbus, S. \& Hawley, J. 1991, \apj, 376, 214

\bibitem[{Balbus (2003)}]{balbusreview:2003}
Balbus, S.~A. 2003, \araa, 41, 555

\bibitem[Bjorkman \& Wood(2001)]{BW:2001} Bjorkman, J.~E., \& 
Wood, K.\ 2001, \apj, 554, 615 

\bibitem[{{Bockel{\' e}e-Morvan} {et~al.}(2002){Bockel{\' e}e-Morvan},
  {Gautier}, {Hersant}, {Hur{\' e}}, \& {Robert}}]{bockmorvan:2002}
{Bockel{\' e}e-Morvan}, D., {Gautier}, D., {Hersant}, F., {Hur{\' e}}, J.-M.,
  \& {Robert}, F. 2002, \aap, 384, 1107

\bibitem[{{Carballido} {et~al.}(2005){Carballido}, {Stone}, \&
  {Pringle}}]{carballido:2005}
{Carballido}, A., {Stone}, J.~M., \& {Pringle}, J.~E. 2005, \mnras, 358, 1055

\bibitem[Chiang \& Goldreich(1997)]{cg97} Chiang, E.~I., \& 
Goldreich, P.\ 1997, \apj, 490, 368 

\bibitem[{{Clarke} \& {Pringle}(1988)}]{clarkepringle:1988}
{Clarke}, C.~J. \& {Pringle}, J.~E. 1988, \mnras, 235, 365

\bibitem[{{Dorschner} {et~al.}(1995){Dorschner}, {Begemann}, {Henning},
  {J\"ager}, \& {Mutschke}}]{dorschner:1995}
{Dorschner}, J., {Begemann}, B., {Henning}, T., {J\"ager}, C., \& {Mutschke},
  H. 1995, \aap, 300, 503

\bibitem[{{Dullemond} {et~al.}(2006){Dullemond}, {Apai}, \&
  {Walch}}]{dullapaiwalch:2006}
{Dullemond}, C.~P., {Apai}, D., \& {Walch}, S. 2006, \apjl, 640, L67

\bibitem[{{Dullemond} \& {Dominik}(2004)}]{duldomdisk:2004}
{Dullemond}, C.~P. \& {Dominik}, C. 2004, \aap, 417, 159

\bibitem[{{Gail}(2001)}]{gail:2001}
{Gail}, H.-P. 2001, \aap, 378, 192

\bibitem[{{Gammie}(1996)}]{gammie:1996}
{Gammie}, C. F. 1996, \apj, 457, 355

\bibitem[{{Gibbard} {et~al.}(1997){Gibbard}, {Levy}, \&
  {Morfill}}]{gibbardlevymorfill:1997}
{Gibbard}, S.~G., {Levy}, E.~H., \& {Morfill}, G.~E. 1997, Icarus, 130, 517

\bibitem[{{Harker} \& {Desch}(2002)}]{harkerdesch:2002}
{Harker}, D.~E. \& {Desch}, S.~J. 2002, \apjl, 565, L109

\bibitem[{{Ilgner} \& {Nelson}(2006)}]{ilgnernelson:2006a}
{Ilgner}, M. \& {Nelson}, R.~P. 2006, \aap, 445, 205

\bibitem[{{J\"ager} {et~al.}(1998){J\"ager}, {Molster}, {Dorschner}, {Henning},
  {Mutschke}, \& {Waters}}]{jaeger:1998}
{J\"ager}, C., {Molster}, F.~J., {Dorschner}, J., {Henning}, T., {Mutschke},
  H., \& {Waters}, L. B. F.~M. 1998, \aap, 339, 904

\bibitem[{{Johansen} \& {Klahr}(2005)}]{johansenklahr:2005}
{Johansen}, A. \& {Klahr}, H. 2005, \apj, 634, 1353

\bibitem[{{Johansen} {et~al.}(2006){Johansen}, {Klahr}, \&
  {Mee}}]{johansenklahrmee:2006}
{Johansen}, A., {Klahr}, H., \& {Mee}, A.~J. 2006, \mnras, 370, L71

\bibitem[{{Keller} \& {Gail}(2004)}]{keller:2004}
{Keller}, Ch. \& {Gail}, H.-P. 2004, \aap, 415, 1177

\bibitem[{{Klerner} \& {Palme}(2000)}]{klernerpalme:2000}
{Klerner}, S. \& {Palme}, H. 2000, Meteoritics \& Planetary Science, vol.~35,
  Supplement, p.A89, 35, 89

\bibitem[{Leinert} {et~al.}(2004)]{leinertvanboekel:2004}
{Leinert}, C., {van Boekel}, R., {Waters}, L.~B.~F.~M., {et al.}
2004, \aap, 423, 537

\bibitem[Lucy(1999)]{Lucy:1999} Lucy, L.~B.\ 1999, \aap, 344, 282 

\bibitem[{{Lynden-Bell} \& {Pringle}(1974)}]{lyndenpring:1974}
{Lynden-Bell}, D. \& {Pringle}, J.~E. 1974, \mnras, 168, 603

\bibitem[{McComb(1990)}]{mccomb:1990}
McComb, W.~D. 1990, The Physics of Fluid Turbulence (Oxford Science
  Publications)

\bibitem[{{Min} {et~al.}(2005){Min}, {Hovenier}, \& {de Koter}}]{minhove:2005}
{Min}, M., {Hovenier}, J.~W., \& {de Koter}, A. 2005, \aap, 432, 909

\bibitem[{{Morfill} \& {Voelk}(1984)}]{morfillvoelk:1984}
{Morfill}, G.~E. \& {Voelk}, H.~J. 1984, \apj, 287, 371

\bibitem[{Natta} {et~al.}(2007)]{nattappv:2007}
{Natta}, A., {Testi}, L., {Calvet}, N., {Henning}, Th.,
{Waters}, R., {Wilner}, D., Protostars and Planets V, 
University of Arizona Press, Tuscon, 2007, p.767

\bibitem[{Oishi}, {Mac Low}, {Menou}(2007)]{oishi:2007}
{Oishi}, J.S., {Mac Low}, M.-M., {Menou}, K. 2007, astro.ph..25490

\bibitem[Pringle(1981)]{pringle:1981} Pringle, J.~E.\ 1981, 
\araa, 19, 137 

\bibitem[{Regev} \& {Gitelman}(2002)]{regev:2002}
{Regev}, O. \& {Gitelman}, L. 2002, \aap, 396 623

\bibitem[{{R{\"u}diger} {et~al.}(2005){R{\"u}diger}, {Egorov}, \&
  {Ziegler}}]{ruedigeregorov:2005}
{R{\"u}diger}, G., {Egorov}, P., \& {Ziegler}, U. 2005, Astronomische
  Nachrichten, 326, 315

\bibitem[{{Semenov} {et~al.}(2004){Semenov}, {Wiebe}, \&
  {Henning}}]{semenovwiebe:2004}
{Semenov}, D., {Wiebe}, D., \& {Henning}, T. 2004, \aap, 417, 93

\bibitem[{{Semenov} {et~al.}(2006){Semenov}, {Wiebe}, \&
  {Henning}}]{semenovmix:2006}
{Semenov}, D., {Wiebe}, D., \& {Henning}, T. 2006, \apjl, 647, L57

\bibitem[{{Servoin} \& {Piriou}(1973)}]{servoin:1973}
{Servoin}, J.~L. \& {Piriou}, B. 1973, phys. stat. sol., 55, 677

\bibitem[{Shakura \& Sunyaev(1973)}]{shaksuny:1973}
Shakura, N.~I. \& Sunyaev, R.~A. 1973, \aap, 24, 337

\bibitem[{{Stone} \& {Balbus}(1996)}]{StoneBalbus:1996}
{Stone}, J.~M. \& {Balbus}, S.~A. 1996, \apj, 464, 364+

\bibitem[{Tscharnuter} \& {Gail}(2007)]{tscharnuter:2007}
{Tscharnuter}, W. M. \& {Gail} H.-P. 2007, \aap, 463, 369

\bibitem[{Tennekes \& Lumley(1972)}]{tennekeslumley:1972}
Tennekes, H. \& Lumley, J.~L. 1972, First course in turbulence (Cambridge: MIT
  Press)

\bibitem[{Trieloff \& Palme(2006)}]{trieloffpalme:2006}
Trieloff, M. \& Palme, H. 2006, in Planet Formation, ed. H.~{Klahr} \&
  W.~{Brandner} (Cambridge, UK: Cambridge University Press)

\bibitem[{{Tutukov} \& {Pavlyuchenkov} (2004)}]{tutukovPav:2004}
{Tutukov}, A.~V. \& {Pavlyuchenkov}, Y.~N. 2004, Astronomy Reports, 48, 800

\bibitem[{{Voelk} {et~al.}(1980){Voelk}, {Jones}, {Morfill}, \&
  {Roeser}}]{voelkmorroejon:1980}
{Voelk}, H.~J., {Jones}, F.~C., {Morfill}, G.~E., \& {Roeser}, S. 1980, \aap,
  85, 316

\bibitem[{Urpin}(1984)]{urpin:1984}
{Urpin}, V.A. 1984, Astron. Zh., 61, 84

\bibitem[{{Zolensky} {et~al.}(2006){Zolensky}, {Zega}, {Yano}, {Wirick},
  {Westphal}, {Weisberg}, {Weber}, {Warren}, {Velbel}, \&
  {Tsuchiyama}}]{zolensky:2006}
{Zolensky}, M.~E., {Zega}, T.~J., {Yano}, H., {Wirick}, S., {Westphal}, A.~J.,
  {Weisberg}, M.~K., {Weber}, I., {Warren}, J.~L., {Velbel}, M.~A., \&
  {Tsuchiyama}, A. e.~a. 2006, Science, 314, 1735

\end{thebibliography}

\end{document}